\documentclass[preprint,5p,twocolumn,11pt,sort&compress]{elsarticle}
\usepackage[english]{babel}
\usepackage{graphicx}
\usepackage[usenames,dvipsnames,svgnames,table]{xcolor}
\usepackage{commath}
\usepackage{bm}
\usepackage{amsfonts}
\usepackage{amsmath}
\usepackage{xcolor}
\usepackage{cuted}
\usepackage{comment}

\usepackage[colorlinks = true,
            linkcolor = blue,
            urlcolor  = blue,
            citecolor = blue,
            anchorcolor = blue]{hyperref}

\def\wde{w}
\def\wgr{w_{\rm gr}}
\def\wbg{w_{\rm bg}}
\def\k{\bm{k}}
\def\d{{\rm d}}
\def\om{\Omega_{\mathrm{m}}}
\def\omo{\Omega_{\mathrm{m},0}}
\def\fsig{f\sigma_8}
\def\sig{\sigma_8}

\makeatletter
\def\ps@pprintTitle{%
  \let\@oddhead\@empty
  \let\@evenhead\@empty
  \let\@oddfoot\@empty
  \let\@evenfoot\@oddfoot
}
\makeatother

\begin{document}

\title{{\bf Measuring dark energy with expansion and growth}}
\date{}

\author[label1]{Louis Perenon \corref{cor}}
\cortext[cor]{Corresponding author: perenon.louis@yahoo.fr}
\address[label1]{Department of Physics \& Astronomy, University of the Western Cape, Cape Town 7535, South Africa}

\author[label2]{Matteo Martinelli}
\address[label2]{INAF -- Istituto Nazionale di Astrofisica, Osservatorio Astronomico di Roma, 00040 Monteporzio Catone, Italy}

\author[label1,label3,label4]{Roy Maartens}
\address[label3]{Institute of Cosmology \& Gravitation, University of Portsmouth, Portsmouth PO1 3FX, UK}
\address[label4]{National Institute for Theoretical and Computational Sciences (NITheCS), Cape Town 7535, South Africa}

\author[label5,label6,label7,label1]{Stefano Camera}
\address[label5]{Dipartimento di Fisica, Universit\`a degli Studi di Torino, 10125 Torino, Italy}
\address[label6]{INFN -- Istituto Nazionale di Fisica Nucleare, Sezione di Torino, 10125 Torino, Italy}
\address[label7]{INAF -- Istituto Nazionale di Astrofisica, Osservatorio Astrofisico di Torino, 10025 Pino Torinese, Italy}

\author[label8,label1]{Chris Clarkson}
\address[label8]{School of Physical \& Chemical Sciences, Queen Mary University of London, London E1 4NS, UK}

\begin{abstract}
We combine cosmic chronometer and growth of structure data to derive the redshift evolution of the dark energy equation of state $w$, using a novel agnostic approach. The background and perturbation equations lead to two expressions for $w$, one purely background-based and the other relying also on the growth rate of large-scale structure. We compare the features and performance of the growth-based $w$ to the background $w$, using Gaussian Processes for the reconstructions. We find that current data is not precise enough for robust reconstruction of the two forms of $w$. By using mock data expected from next-generation surveys, we show that the reconstructions will be robust enough and that the growth-based $w$ will out-perform the background $w$. Furthermore, any disagreement between the two forms of $w$ will provide a new test for deviations from the standard model of cosmology.
\end{abstract}


\maketitle

\section{Introduction}

Understanding dark energy remains a key problem in cosmology. A key quantity to trace the behaviour of dark energy is its equation of state $\wde(z)=\bar p_{\rm de}(z)/\bar\rho_{\rm de}(z)$, the ratio of the (spatially averaged) pressure and energy density. Deriving $w$ is typically done using observations which infer background quantities such as the distance-redshift relation or the expansion rate as a function of redshift, $H(z)$. Then $w$ can be estimated using a simple parametrisation or via non-parametric approaches (see e.g. \cite{Sahni:2006pa, Silvestri:2009hh, Huterer:2017buf} for reviews). A key problem in this approach is that whatever data is used, it needs to be differentiated in some way. In the case of distance data, such as from type Ia supernovae, two derivatives are required, greatly enhancing the error on $w(z)$ and increasing the sensitivity to the reconstruction approach that is used. While direct measurements of $H(z)$ only require one derivative to be taken, the observations themselves are much more sparse, e.g. from cosmic chronometers, or require extra assumptions, e.g. from baryon acoustic oscillations (BAO). 

It is therefore important to access $w(z)$ in as many agnostic ways as possible. Here we present a new approach to direct reconstructions of $w(z)$ by including measurements of the growth rate $f(z)$. We show that in principle $w(z)$ can be found if we know $f(z)$ and $f'(z)$, together with $H(z)$ (and a prior on $\Omega_{m,0}$). Smoothing the measurements of these functions can give a direct measurement of $w(z)$. We do this smoothing using Gaussian Processes (GP), which has been shown to be particularly stable for taking derivatives of raw data~\cite{Seikel:2012uu,Seikel:2013fda}. 
This approach gives a new method for probing dark energy and gravity. In addition, it provides an important consistency test: if the growth-based $w$ disagrees with the purely background $w$, this would signal a breakdown of the simple effective fluid model of dark energy (including quintessence models). The mismatch would be indicative of a clustering or interacting form of dark energy -- or of a modification of general relativity. 

The paper is organised as follows. We show in \autoref{sec:eos} how the dark energy equation of state can be obtained from background data only and derive how growth of structure data can be brought into play. In \autoref{sec:current}, we compare the predictions of the equations of state using current data, while in \autoref{sec:forecasts} we analyse their performances using forecasts on mock data and present the diagnostic which arises from comparing the background and growth equations of state. We summarise our findings in \autoref{sec:discussion}.

\section{Dark energy equation of state}\label{sec:eos}

Here we present two different methods to reconstruct $w(z)$ directly from data. The first is well known, and uses data that measures the background Hubble rate. The second is new and uses the time evolution of the growth rate. 

The standard simple model of dynamical dark energy is defined by its equation of state $w(z)$. The dark energy is assumed to be non-interacting, so that it obeys energy conservation: $\dot{\bar{\rho}}_{\rm de}+3(1+w)H\,{\bar{\rho}}_{\rm de}=0$ (see e.g. \cite{Kase:2020hst} on interacting dark energy). It is assumed to be non-clustering on observable scales, which entails an implicit assumption that its speed of sound is close to 1 (see e.g. \cite{Hassani:2020buk,Batista:2021uhb} on clustering dark energy).

The evolution of the Hubble rate is given in general relativity on a spatially flat background by
\begin{align}\label{doth}
{\d \ln H \over \d \ln (1+z)}
&\equiv -\frac{\dot H}{H^2}=
 {3\over 2}+{3\over 2}\,\left(1-\Omega_{\rm m}\right)\,w\;,\\
\label{eq:omh}
 \om(z) &= \omo\,H_0^2\, \frac{(1+z)^3}{H^2(z)}\;.
\end{align}
We can then reconstruct the equation of state using \autoref{doth}:
\begin{equation}\label{eq:wbg}
w_{\rm bg} = \dfrac{1}{1 - \om}\, \left[\dfrac23\, \dfrac{\d\ln H}{\d\ln (1+z)} -1 \right], 
\end{equation}
where the subscript `bg' indicates that this reconstruction involves only background quantities, i.e.\ $H(z)$, $\d H(z)/\d z$ and $\omo$.

However, it also is possible to infer the equation of state from observations of cosmological perturbations. For example, redshift-space distortions (RSD) in galaxy correlations aim to measure the linear growth rate $f$ of large-scale structure, a powerful probe of gravity. For standard models of dark energy in general relativity, at late times the linear growth factor $D$ is scale-independent, if we neglect the effects of massive neutrinos. In this case, the linear matter density contrast is separable: $\delta_{\rm m}(z,\k)=D(z)\,\delta_{\rm m}(0,\k)$. It then follows that the growth rate is scale independent:
\begin{equation}\label{eq:fdef}
f\equiv -\frac{\partial\ln \delta_{\rm m} }{\partial \ln (1+z)}
= - \frac{\d\ln D }{\d \ln (1+z)}\;.
\end{equation}

Massive neutrinos introduce suppression and scale dependence in the linear power spectrum, but these effects are significantly smaller for the linear growth rate $f$ \cite{Villaescusa-Navarro:2017mfx}. Current constraints on the total mass of neutrinos are $M_\nu < 90\,\mathrm{meV}$ at 95\% confidence level (CL) \cite{DiValentino:2021hoh}. In this mass range, the scale-dependent suppression of $f$ is at sub-percent level on linear scales \cite{Aviles:2021que}.

The perturbed conservation equations and the Poisson equation imply that
\begin{equation}\label{edel}
\ddot{\delta}_{\rm m} + 2\,H\,\dot{\delta}_{\rm m}- \frac{3}{2}\,\om\,H^2\,\delta_{\rm m} = 0\,.
\end{equation}
By \autoref{eq:fdef}, using $\d z=-(1+z)\,H\,\d t$, we have $\dot{\delta}_{\rm m}=H\,f\,{\delta}_{\rm m}$ and $\ddot{\delta}_{\rm m}=[(H\,f)^{\displaystyle{\cdot}}+(H\,f)^2]\,{\delta}_{\rm m}$. Then \autoref{edel}
can be rewritten as an evolution equation for the growth rate: 
\begin{align}\label{dfdz}
\frac{\d\ln f}{\d\ln(1+z)}
&= f+2 -\frac{\d\ln H}{\d\ln(1+z)} -{3\om\over 2f}
\\
&= f+\frac12 +{3\over2}\,\left(\om-1 \right)\wde-{3\,\om\over 2\,f} \,.\nonumber
\end{align}
Here the first line follows directly from \autoref{edel} and the second line uses \autoref{doth}.

The second line in \autoref{dfdz}, which applies in standard dark energy models, implies that we can reconstruct $w(z)$ in a new way that incorporates structure formation and not only background information: 
\begin{equation}\label{eq:wgr}
w_{\rm gr} = \frac1{1-\om}\,
\left\{ \frac23\,\left[f - {\d\ln f\over \d\ln(1+z)} \right] -\frac{\om}{f} + \frac13 \right\}\,.
\end{equation}
The subscript `gr' indicates that we use growth data $f(z)$ and $\d f(z)/\d z$, as well as background data -- i.e. $H(z)$ data and a prior on $\omo$. This gives a new method to extract the dark energy equation of state. Note that the concept of splitting key cosmological quantities into their background and perturbation or growth counter parts is not novel. More model-dependent methods have been considered in the past \cite{Wang:2007fsa, Ruiz:2014hma, Bernal:2015zom, DES:2020iqt, Ruiz-Zapatero:2021rzl, Andrade:2021njl}. 

In our approach, by considering data on $H$, we can employ GP to reconstruct the evolution of $H$ and then deduce $\d H/\d z$ and $\om$, given a value of $\omo$. Then $\wbg$ can be fully determined. Considering data on $f$ and its GP reconstruction in addition, $\wgr$ can be fully determined also. Then determining consistency between these gives a new diagnostic of dark energy and gravity, which we explore for present and upcoming experiments. 

\section{Reconstructions with current data}\label{sec:current}

\begin{figure}[!]
\begin{center}
\hskip-1.65mm\includegraphics[width=\columnwidth]{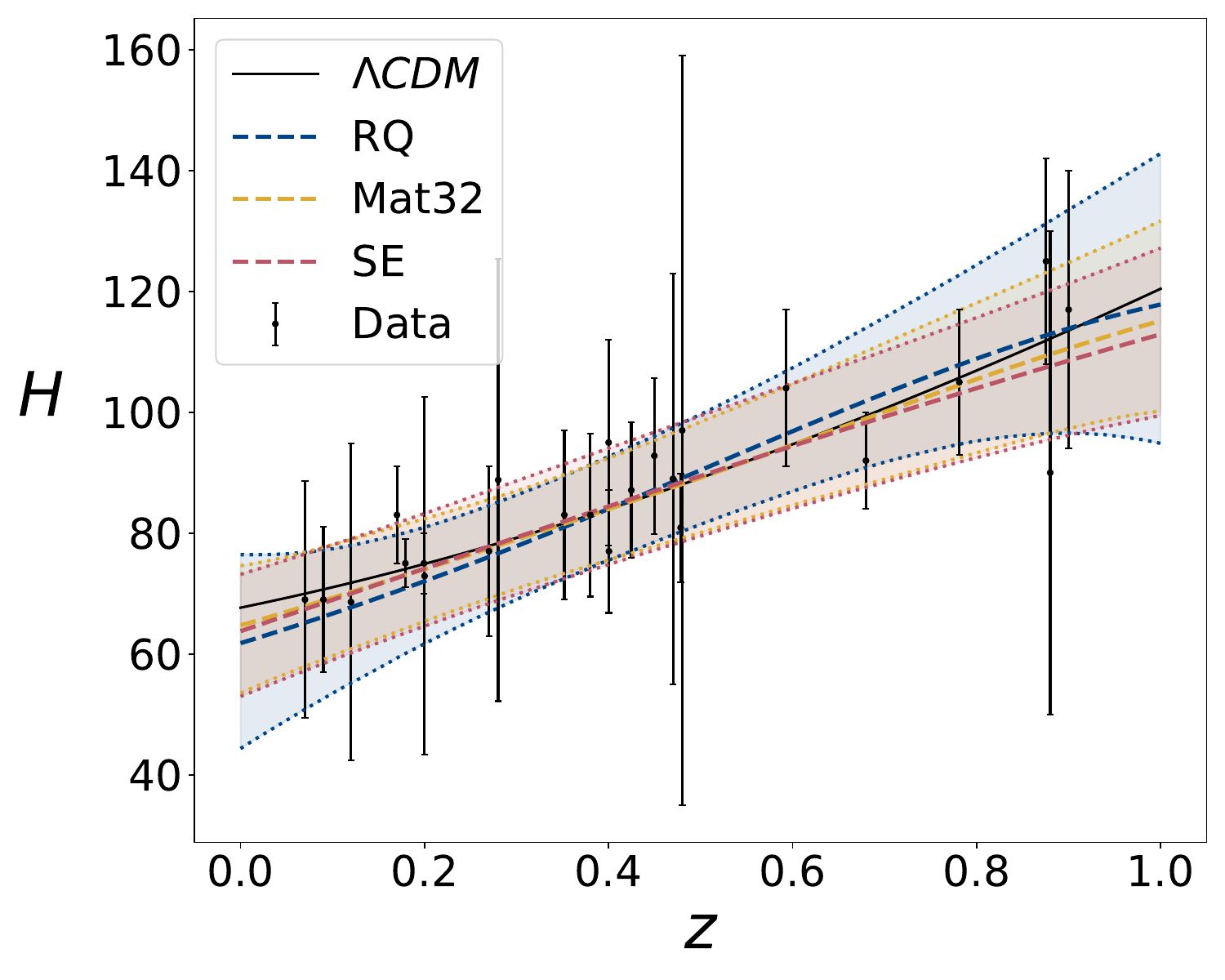}\vskip-10.23mm
\includegraphics[width=0.988\columnwidth]{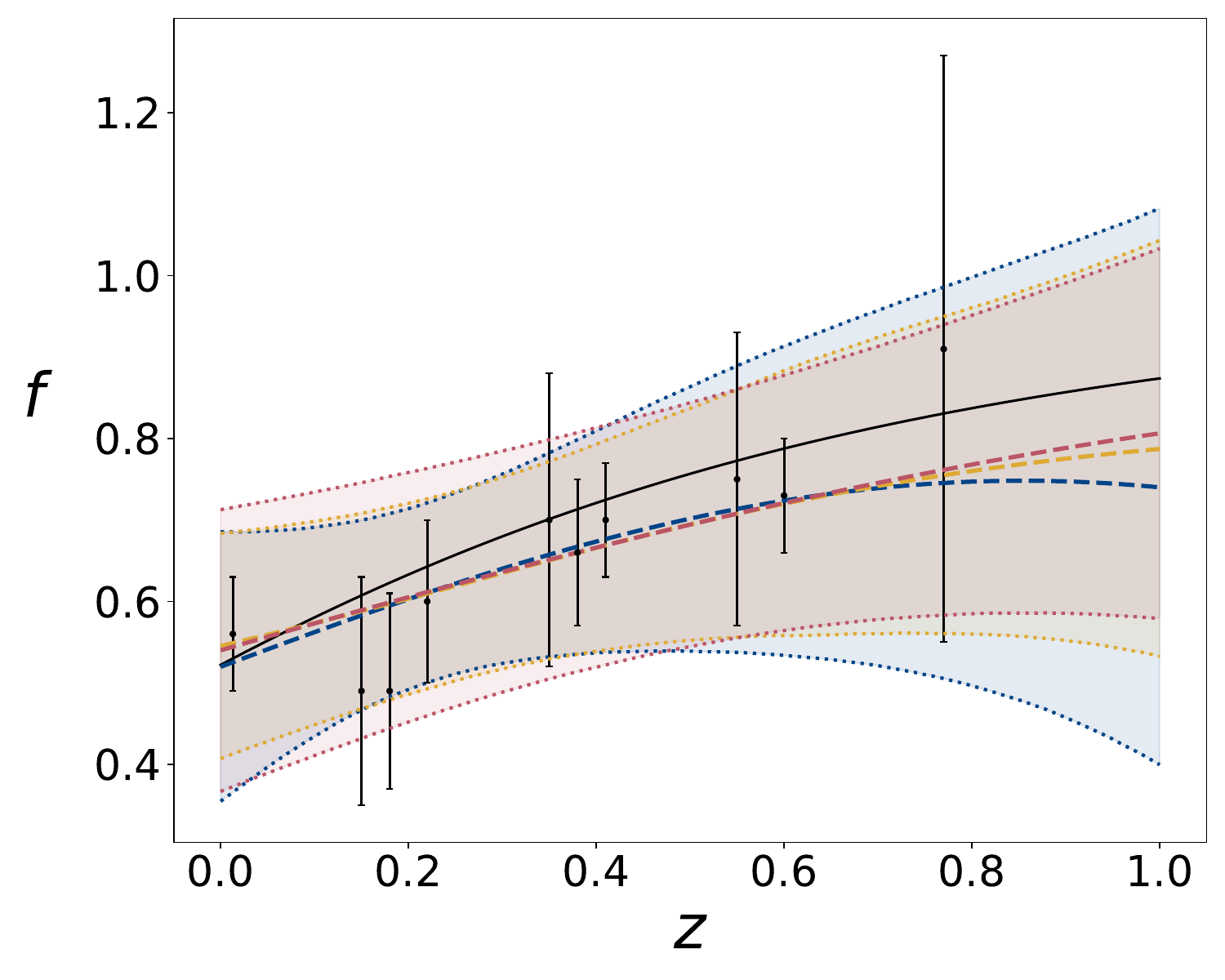}
\caption{Optimised reconstruction of $H$ (top) and $f$ (bottom) using 3 GP kernels. Means are dashed lines, shaded areas between dotted lines are 95\% CL. $\Lambda$CDM is the solid black line.}
\label{fig:current_H_f}
\end{center}
\end{figure}

We use the `Humble Code for Gaussian Processes' that we have made publicly available.\footnote{ \url{https://github.com/louisperenon/HCGP}} This code is able to compute automatically any combinations of kernels and their derivatives since it uses the python package SymPy. The first step to obtain the dark energy equations of state is to reconstruct the functions $H$ and $f$ with GP. We probe how robust the reconstructions are by looking at dependence on priors and assumptions, in order ensure that our conclusions are as agnostic as possible. Robustness tends to apply when the data is numerous and precise, an assumption that might not apply to current data.

For $H(z)$, we use cosmic chronometer (CC) data, via the compilation in Appendix A of \cite{Bernardo:2021cxi}. We choose not to use the BAO data here, since they are less model-independent, requiring a prior on the sound horizon at recombination in order to be converted into measurements of $H(z)$. The $f(z)$ data are from \cite{Avila:2022xad}, a compilation obtained from different tracers. This compilation contains only uncorrelated data and direct measurements of $f(z)$. We restrict all datasets to $z \le 1$ in order to consider only the redshift range where data on both $f$ and $H$ are the most available.  

As a first step, we explore the dependency of our results on the choice of priors for the means of the reconstructed functions. We choose not to enforce any mean prior on $H$ and $f$. However, we also considered the case where predictions on these functions inferred from Planck data (within the $\Lambda$CDM model) are used as mean priors. We find that this produces negligible differences with the null prior case.

Secondly, we examine how the choice of GP kernel impacts the reconstructions. Since $H$ and $f$ are smooth functions of $z$, we use simple stationary kernels, comparing 3 kernels to cover the possible evolution of the data (see \cite{Rasmussen} for details of these kernels): 
\begin{enumerate}
  \item SE (squared exponential) is one of the most commonly used in cosmology as it is simple and predicts the smoothest evolution generally;
  \item Mat32 (Matern $3/2$) can capture the sharpest variations of the Matern class and is best able to capture noisy data;
  \item RQ (rational quadratic) can be seen as an infinite sum of SE kernels with different correlation lengths, and an extra hyperparameter.
\end{enumerate}

We optimise a GP for each kernel choice, for both $H$ and $f$, ensuring that the allowed ranges on kernel hyperparameters are always chosen large enough to yield no effect. We take $[10^{-5},\,10^5]$ for each hyperparameter. The results are displayed in \autoref{fig:current_H_f}. The $\Lambda$CDM model that we use is from CMB+BAO data (the last column of Table 2 in \cite{Planck:2018vyg}). It is recovered within the 95\% confidence regions of all the reconstructions, but some differences are evident. The RQ kernel tends to yield the largest errors at the extremes of the data range, while Mat32 gives the smallest errors where the data clusters. 

The data lacks constraining power. The fact that these three kernels yield different results overall is a hint that the quality of the current data is still too poor to produce more robust reconstructions. By contrast, reconstructions with future galaxy survey data do not suffer from this, see e.g. \cite{Perenon:2021uom}, and we will exploit this in \autoref{sec:forecasts}.

\begin{figure}[!]
\begin{center}
\includegraphics[width=\columnwidth]{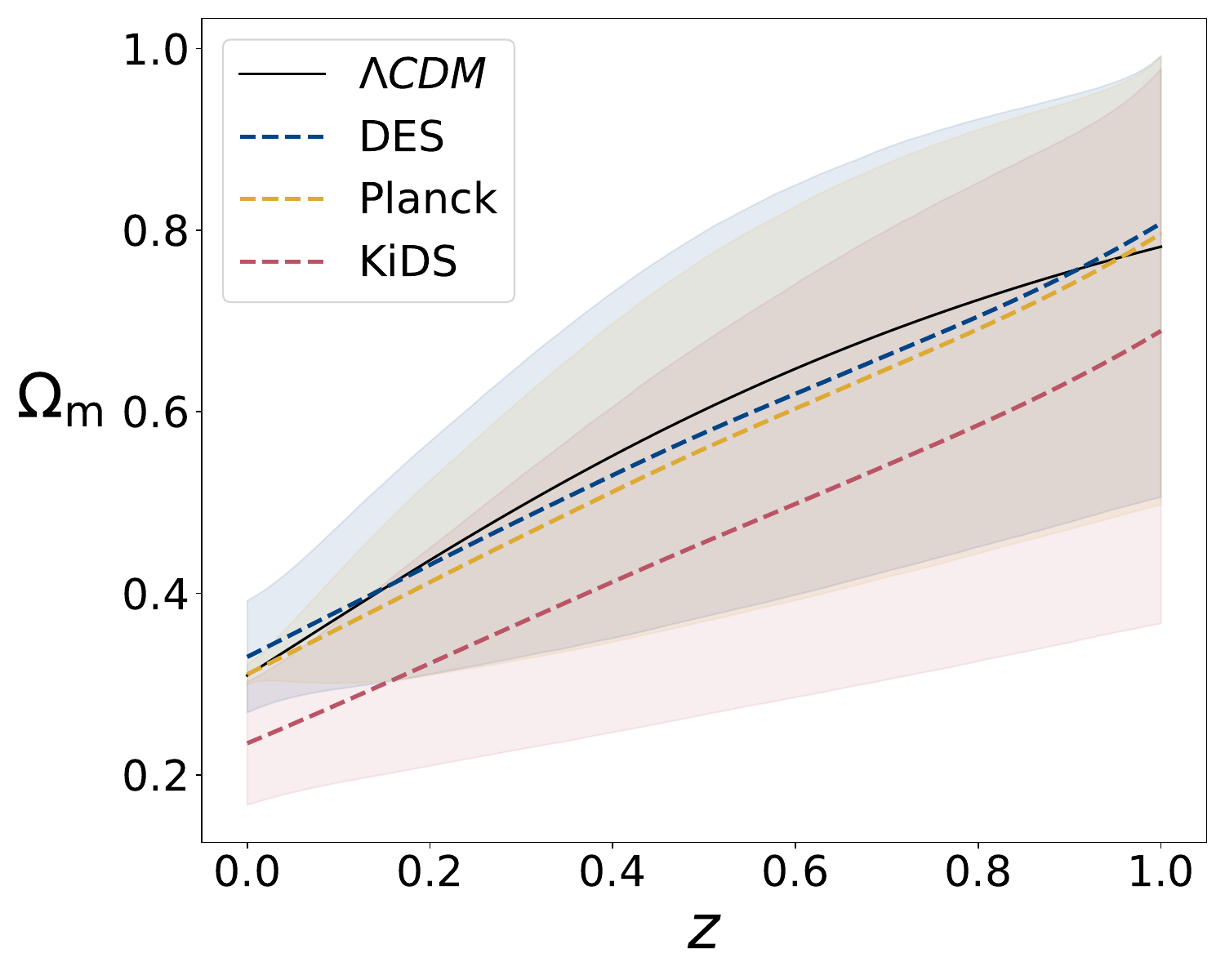}
\caption{Matter density $\om(z)$ from GP reconstruction of $H(z)$, with Mat32 kernel and 3 choices of prior on $\omo$: DES Y3 (weak lensing and galaxy clusters); Planck CMB+BAO; KiDS-DR3 (AMICO galaxy clusters). Means are dashed lines, shaded areas are 95\% CL. }
\label{fig:current_om}
\end{center}
\end{figure}

Another key ingredient in the predictions of $\wbg$ and $\wgr$ is $\om$, given in \autoref{eq:omh}. It must be computed through the Monte Carlo realisation of two quantities, one from the GP on $H(z)$ and one from $\omo$. While the possible realisations of $H(z)$ are entirely dictated by the CC data and the assumptions on the GP, the choice of $\omo$ brings in an extra layer of prior knowledge that we need to examine. To illustrate how different values of $\omo$ can impact the predictions, we use a Mat32 kernel for $f$ and $H$ and consider 3 different priors: 

\begin{enumerate}
    \item the Planck CMB+BAO constraint of Table 2 in \cite{Planck:2018vyg}, $\omo = 0.3111 \pm 0.0056$\,;
    \item the KiDS-DR3 constraint from AMICO galaxy clusters \cite{Lesci:2020qpk},  $\omo = 0.24\,^{+0.03}_{-0.04}$\,;
    \item the DES Y3 constraint from weak lensing and galaxy clusters \cite{DES:2021wwk}, $\omo = 0.339\,_{-0.031}^{+0.032}$\,.
\end{enumerate}

The results on the predictions of $\om(z)$ are shown in \autoref{fig:current_om}. The choice of prior on $\omo$ is a key factor in the precision on $\om(z)$ at small redshifts, as expected. As redshift increases, uncertainties from the different prior choices becomes similar. Indeed, the error on $\om(z)$ becomes dominated by the uncertainty on $H(z)$. 

Furthermore, there is an important caveat to consider in the prediction of $\om(z)$. Given our assumption of a spatially flat Universe, used to obtain \autoref{eq:wbg} and \autoref{eq:wgr}, we have a theoretical prior $\om(z)<1$. Violating this condition is unphysical and would lead to divergence in $w$ due to the $1/(1-\om)$ factors. Given the precision of the reconstruction of $H$ and the choice of prior on $\omo$, some realisations are however likely to violate this condition, in particular at higher redshift when matter starts to dominate over dark energy. The more precise is the data, the less probable are such occurrences. Using the kernel that leads to the smallest errors, i.e. Mat32, we find that with currently available data, 47\%, 18\% and 57\% of the realisations violate such a condition for the 3 choices of $\omo$ prior. This effect is another indication that more precise data is desirable to produce more robust predictions. 

For now, we choose to reject the realisations that do not satisfy $\om (z) < 1$. The effect of this choice artificially pushes the mean of the prediction of $\om$ to lower values relative to the standard model, as seen in \autoref{fig:current_om}. 

\begin{figure}[!]
\begin{center}
\includegraphics[width=\columnwidth]{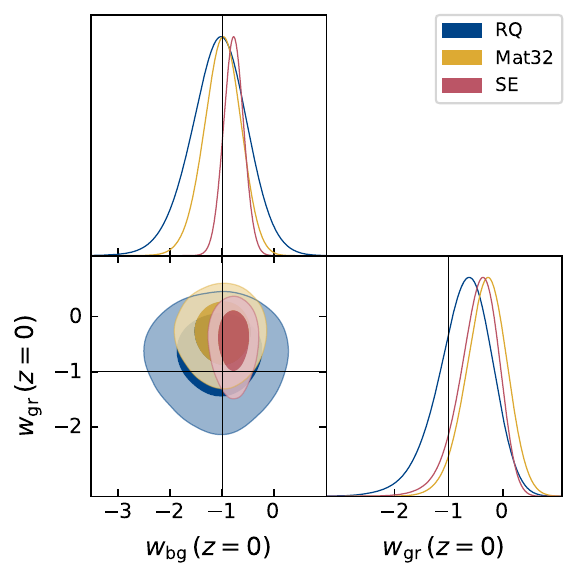}
\includegraphics[width=\columnwidth]{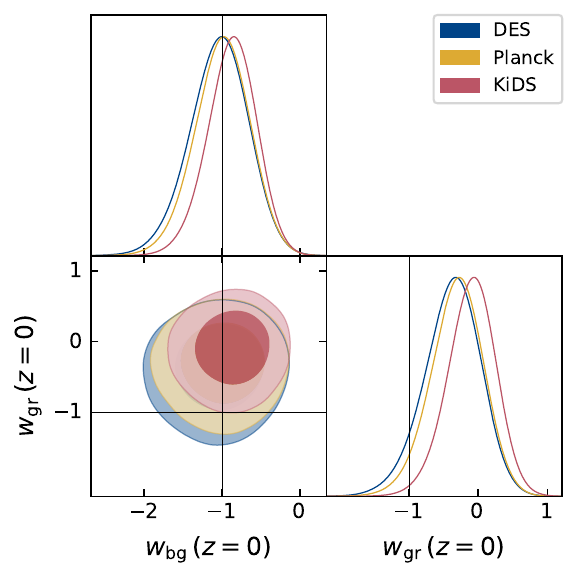}
\caption{Distribution of $\wbg$ and $\wgr$ at $z=0$ from Monte Carlo realisations of GP-generated $f$ and $H$ and the derived $\om$. {\em Top:} For Planck prior on $\omo$ and different GP kernels (see \autoref{fig:current_H_f}). {\em Bottom:} For Mat32 kernel and different priors on $\omo$ (see \autoref{fig:current_om}). Shading levels are 68\% and 95\% CL.}
\label{fig:current_w0}
\end{center}
\end{figure}

To derive the evolution of $\wbg$ and $\wgr$, we use a Monte Carlo procedure. With a GP mean and covariance for $f(z)$ and $H(z)$, we can derive the mean and covariance of their derivatives (see \cite{Seikel:2012uu} for details). Supplementing this with a prior on $\omo$, we derive predictions of $\om(z)$. We then randomly simulate realisations of each of these quantities and from these derive a prediction of $w(z)$ at each step, following \autoref{eq:wbg} and \autoref{eq:wgr}. Their distribution at $z=0$ for the different choices of kernel and priors is given in \autoref{fig:current_w0}. Note that the constraints we find on $\wbg$ are in good agreement with the recent findings of \cite{Kumar:2022mtx} using the same data set but a different method.

In the top panel, we use the Planck prior on $\omo$ and compare the kernels. The RQ kernel leads to much larger errors, showing that the derivation of $w$ can be very sensitive to kernel choice given the data that we use for the reconstruction. Note that the standard model ($\wbg=\wgr=-1$) is consistent within 95\% CL for each kernel. Interestingly, there is a tendency for $\wgr>\wbg$. In addition, the SE and RQ kernels, which yield a $\wbg$ centred around the standard model, produce a $\wgr$ systematically larger than $-1$. This is a consequence of the fact that growth data has a slight preference for suppressed growth relative to the standard model \cite{Kazantzidis:2018rnb, Kazantzidis:2018jtb, Perenon:2019dpc, Benisty:2020kdt}. As a result, the reconstructions of $f$ tend to lie further below the standard model as can be seen in \autoref{fig:current_H_f}. 

In the bottom panel, we fix the kernel of the reconstruction to Mat32, and compare the different priors on $\omo$. The differences induced by changing prior are milder than those induced by changing kernel. We also observe that there is not a one to one correspondence between the error on $\omo$ and that on $w(0)$. For example, the tightest prior on $\omo$ is from Planck, and it does not lead to the smallest error. It is in fact the error on the derivative of $H$ which plays a key role. This is another example showing that the details of the GP reconstructions strongly affect the derivation of $w$ and hence the choice of kernel is more important than the choice of $\omo$ prior. 

We observe a systematic increase of $w(0)$ when $\omo$ decreases, although the results are always compatible with each other. This arises since a decrease in $\omo$ increases the factor $(1-\omo)^{-1}$ in \autoref{eq:wbg} and \autoref{eq:wgr}. The increase is enhanced for $\wgr(0)$ by the term $-\omo/f(0)$. Nevertheless, the tendency to have $\wgr>-1$ is present no matter what the choice of $\omo$ prior is.

\begin{figure}[!]
\begin{center}
\includegraphics[width=\columnwidth]{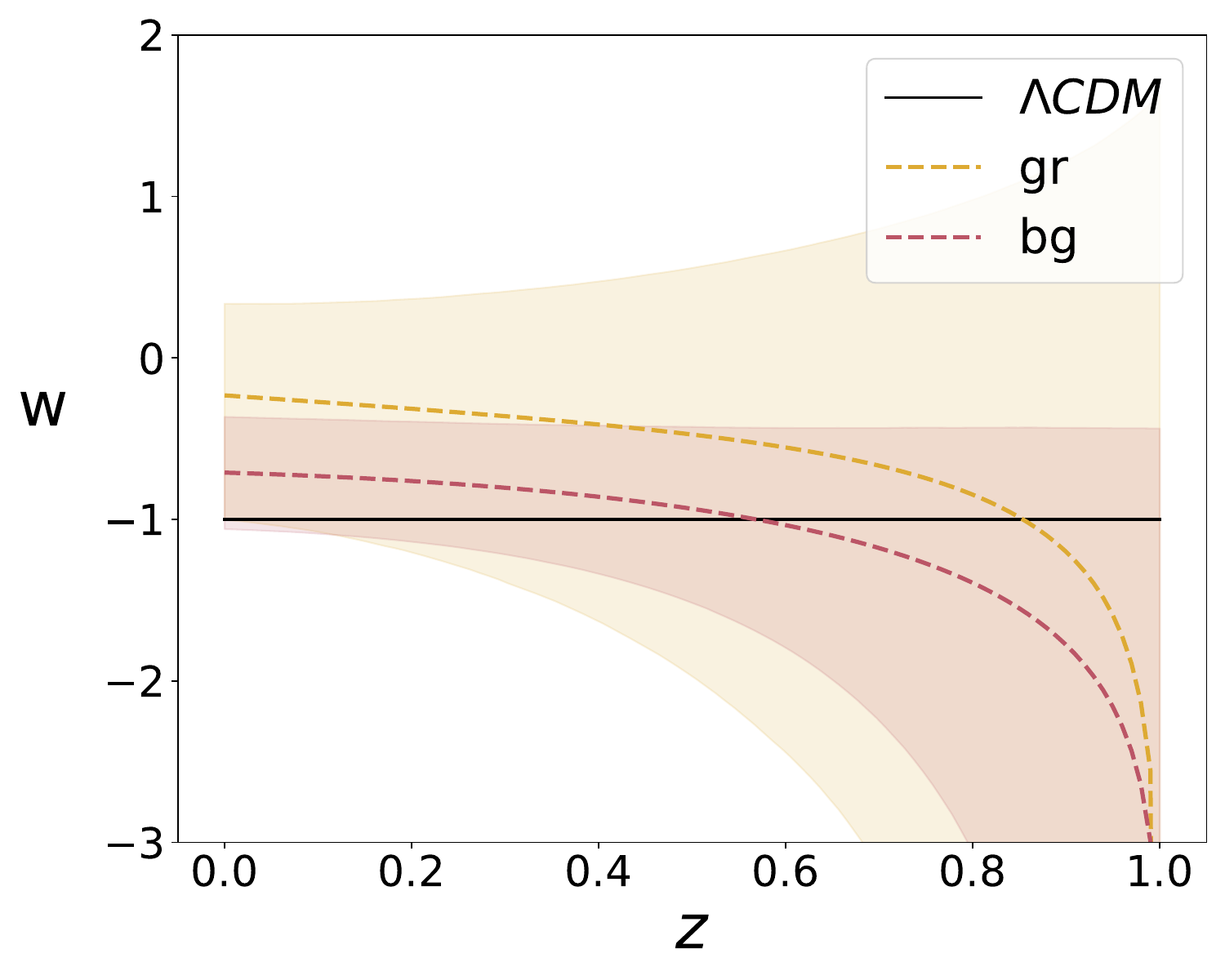}
\caption{Derived predictions of $\wbg$ and $\wgr$ from Monte Carlo realisations of GP-generated $f$ and $H$, with Mat32 kernel and KiDS prior on $\omo$.}
\label{fig:current_w}
\end{center}
\end{figure}

Using a Mat32 kernel and the KiDS prior on $\omo$, we display the redshift evolution of $\wbg$ and $\wgr$ in \autoref{fig:current_w}. The downward trend of the means at higher redshifts is a result of the $\om (z) < 1$ prior. However, as expected we find that such a requirement does not affect the distribution of $w$ at low redshifts. 

Although the reconstructed $\wbg$ and $\wgr$ are compatible with the standard model within $2\sigma$, the tendency of the perturbation data to prefer a lower growth rate than in the standard model leads to hints for $\wgr$ different from $\wbg$, with $\wgr > -1$. This could be a signal of deviations from the standard model. However, the precision and number of current data is too poor to make robust claims. The predictions are too sensitive to the details of the reconstructions, i.e. the GP kernels and the prior on $\omo$. The analysis we developed here should become much more stringent with future data. 

\section{Forecasts with future surveys} \label{sec:forecasts}

\begin{figure}[!]
\begin{center}
\includegraphics[width=\columnwidth]{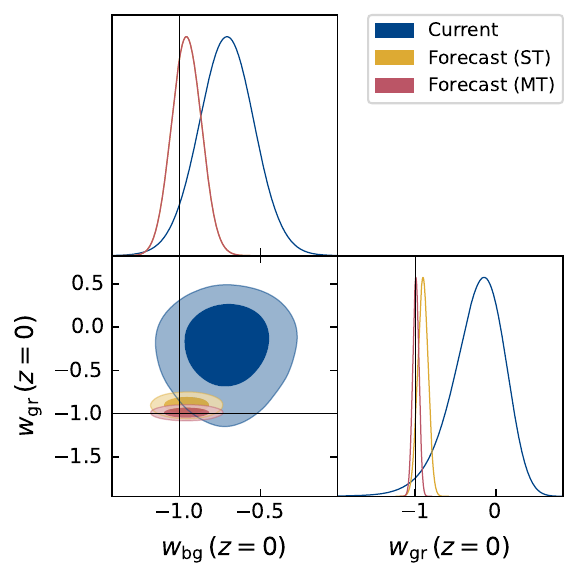}
\includegraphics[width=\columnwidth]{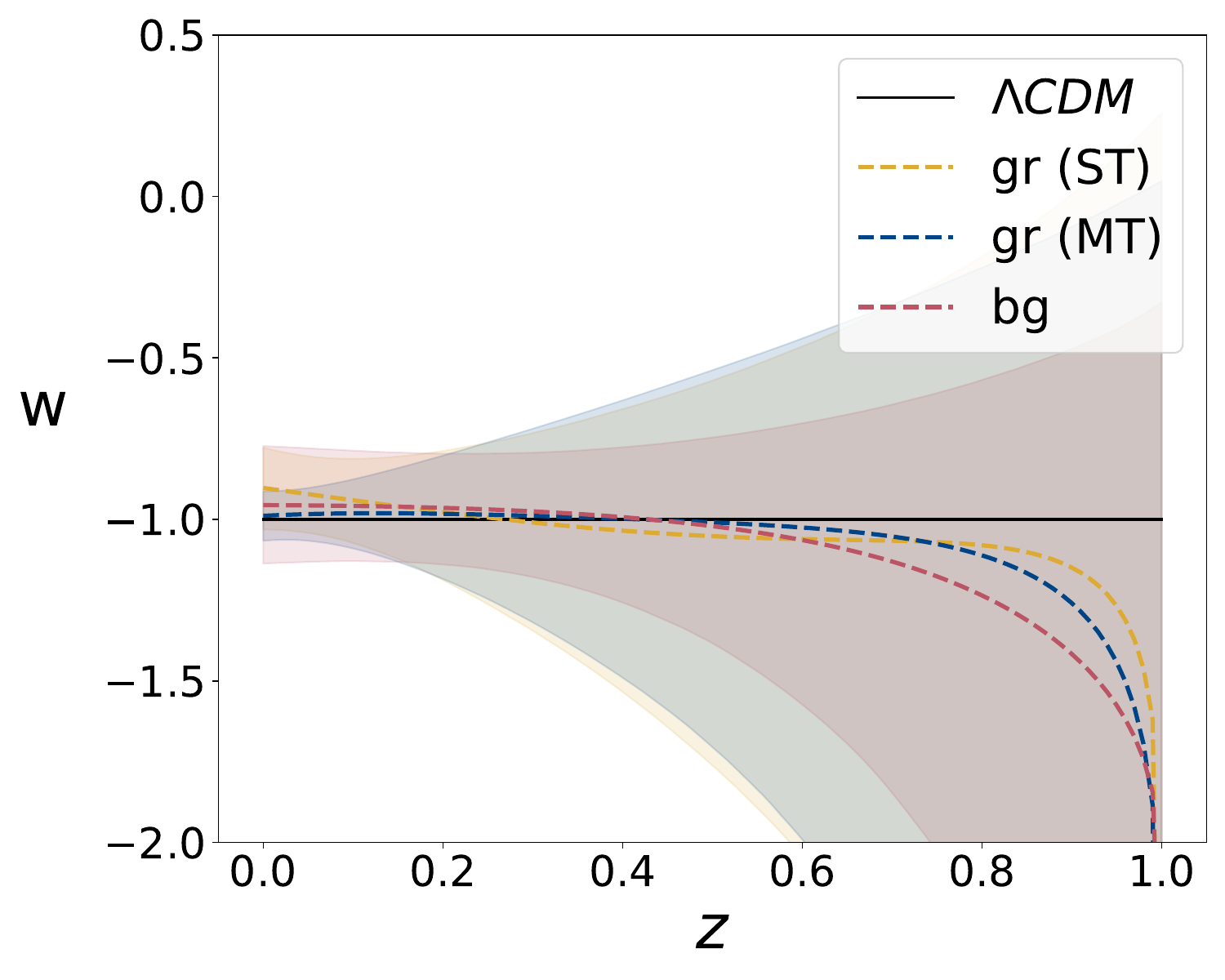}
\caption{Predictions of $\wbg$ and $\wgr$ from Monte Carlo procedure based on single- and multi-task GP reconstructions of $f$ and $H$ with mock future data. \emph{Top:} Distributions at $z=0$, compared to current data (as in \autoref{fig:current_w} bottom panel, with KiDs prior). \emph{Bottom:} Redshift evolution of $\wbg$ and $\wgr$.}
\label{fig:forecast}
\end{center}
\end{figure}

In this section we estimate the precision that could be achieved by future surveys on the $w$ derivations. We also analyse how effective this can be as a test of the standard model. 

\subsection{Reconstructing $\wbg$ and $\wgr$}
\vskip1mm

Instead of choosing specific surveys, we simply assume nominal surveys that deliver percent-level precision on $H$ and $f$ at low $z$, with errors growing as $z$ increases. We create mock datasets for $f(z)$ and $H(z)$ with $\Lambda$CDM as fiducial. This gives predicted values at 11 redshifts uniformly distributed between redshift 0 and 1.

For $f$, we assume the errors start at 1\% and grow as $(1+z)$. For the cosmic chronometer data on $H$ we use the prescription of \cite{Martins:2016bbi}, where the error on the spectroscopic velocity increases linearly from $1\%$ to {$11\%$ from the first to the last measurement.} Then for $f$ and $H$, we draw the mocks from multivariate Gaussian distributions with mean given by their fiducial prediction. The diagonal covariance matrix entries are their errors squared. We also choose a prior on $\omo$ with $1\%$ error, in line with the expected precision of future weak lensing surveys. 

We find that this precision on the data yields GP reconstructions much less dependent on the choices of kernel and priors than for current data. The reconstructions are now completely driven by the data \cite{Perenon:2021uom}. We therefore choose an SE kernel and a null mean prior, as in \cite{Perenon:2021uom}. We find that only 2.7\% of the reconstructions are rejected by the $\om (z ) < 1$ condition. The reconstructions produced are now much more robust and unbiased, compared to those extracted from current data.

At this level of precision, the benefits of including growth data to reconstruct $w$ become apparent. We comment first on the results at $z=0$, shown in \autoref{fig:forecast}, top panel. Comparing the widths of the constraints, it is evident that the $\wbg$ error which uses only future cosmic chronometer data is decreased by factor $\sim2$ relative to current data. When including future growth data, the $\wgr$ error decreases by more than a factor of 5. We obtain for the forecast using background data:
\begin{equation}
w_{\rm bg} (0) = -0.956\,_{-0.181}^{+0.183} \qquad (95\%\,{\rm CL}). 
\end{equation}
This is improved by 44\% when adding the forecast growth data: 
\begin{equation}\label{wgrst}
w_{\rm gr} (0) = -0.903\,_{-0.128}^{+0.125} \qquad (95\%\,{\rm CL}). 
\end{equation}

Upcoming surveys will not only improve measurements of the growth rate $f(z)$, but will also obtain precise estimate of the two related functions, $\sigma_8(z)$ and $(f\sigma_8)(z)$. Spectroscopic galaxy surveys will allow for precise measurements of $(\fsig)(z)$, while combinations of galaxy-galaxy lensing with RSD \cite{delaTorre:2016rxm, Shi:2017qpr, Jullo:2019lgq} or combining matter power spectrum and bispectrum \cite{Gil-Marin:2016wya, Ivanov:2021kcd, Philcox:2021kcw, DAmico:2022gki, Cabass:2022ymb}, will break the $\fsig$ degeneracy and extract measurements of $f$ and $\sig$. 

With these measurements in hand, one can avoid reconstructing only $f$ by combining the information contained in the measurements of the three functions. This can be done by reconstructing simultaneously $f$, $\sig$ and $\fsig$ via the use of multi-task GP, which greatly improves the error on the prediction of each quantity \cite{Perenon:2021uom}. Following the procedure of \cite{Perenon:2021uom}, we use mock data to forecast $\sig$ and $\fsig$, using the same prescriptions as we have for $f$, and thus reconstruct simultaneously the three functions. Using now this multi-task (MT) reconstruction of $f$ in the derivation of $\wgr$, we obtain 
\begin{equation}
w_{\rm gr, MT}(0) = -0.989\,_{-0.077}^{+0.076} \qquad (95\%\,{\rm CL}). 
\end{equation}
This yields a 67\% improvement on the single-task (ST) prediction \autoref{wgrst}. 

At higher redshifts the picture is quite different. The error on $w$ increases with $z$, since the error on the forecast data does. However, \autoref{fig:forecast} bottom panel shows that the error on $\wbg$ is smaller than that on $\wgr$ -- since $\wgr$ depends on $\om/f$. As the error on $\om$ increases with $z$, it comes to dominate the overall error on $\wgr$. For the same reason, there is virtually no gain at larger redshifts made by considering a multi-task reconstruction, even though the precision of the reconstruction on $f$ is improved at all redshifts relative to the single-task case. 

In order to illustrate the previous findings further, we run a large number of realisations of $w$, assuming that {$\om(z)$} is known perfectly, i.e. fixing it to the the fiducial. The results are informative: the error on $\wbg$ does not improve while that on $\wgr$ does considerably and is smaller at almost all redshifts. Its multi-task version is a factor $~2$ more precise than the single-task case. This shows that the error on $\om$ is significantly more detrimental to the prediction of $\wgr$, due to the $\om/f$ term. At low redshifts, the error on $\wgr$ is always lower than for $\wbg$. For $\wgr$ to become a more competitive diagnostic at high $z$, more precise constraints on $\om(z)$ are required. 
\begin{figure}[!ht]
\begin{center}
\includegraphics[width=\columnwidth]{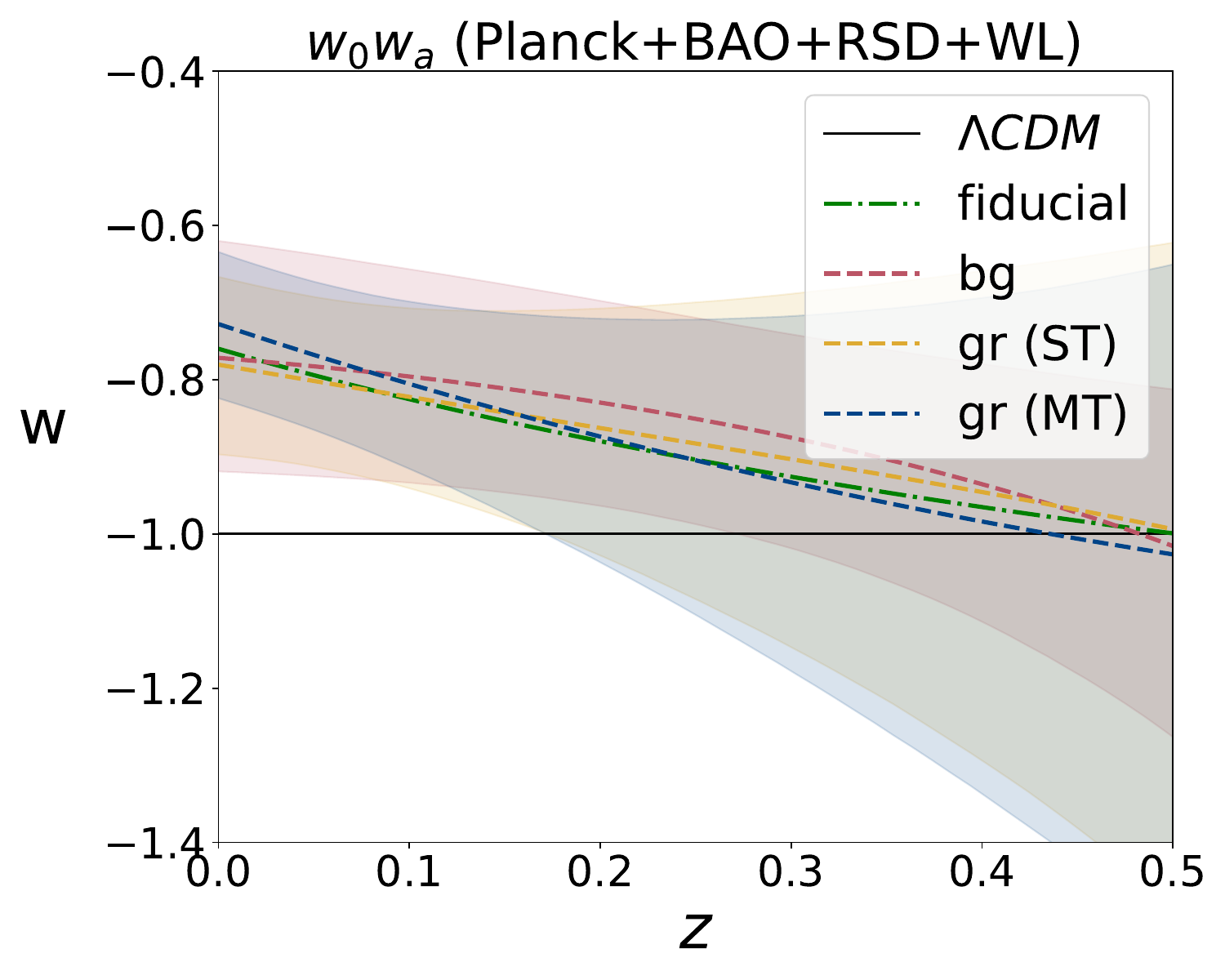}
\includegraphics[width=\columnwidth]{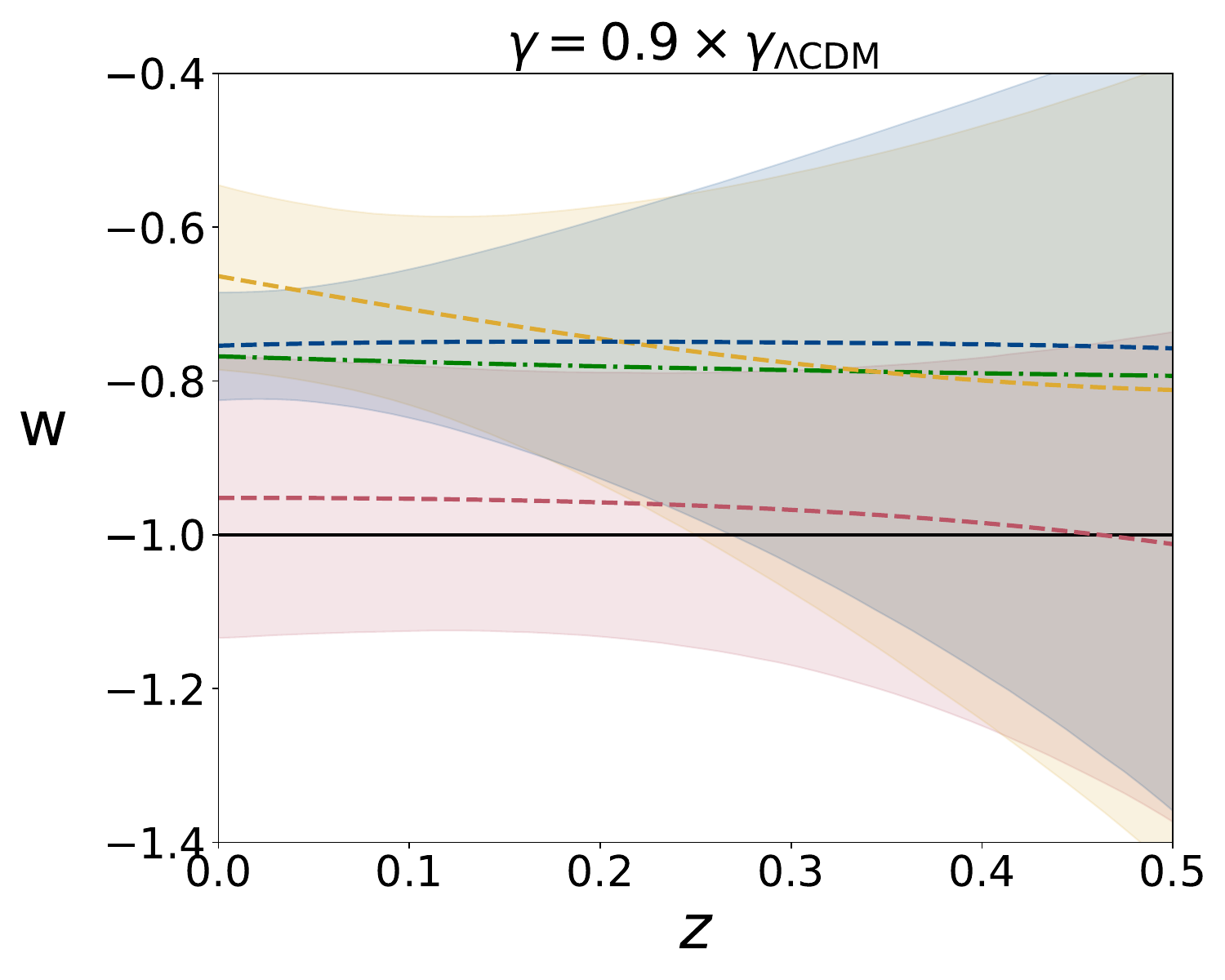}
\caption{As in \autoref{fig:forecast} (bottom)
but using mock data from non-$\Lambda$CDM models. \emph{Top:} Predictions for $w_0w_a$CDM as constrained by Planck \cite{Planck:2018vyg}. \emph{Bottom:} Predictions for a model with $\Lambda$CDM fiducial on $H$, while $f=\om^\gamma$ with $\gamma\neq\gamma_{\lambda{\rm CDM}}=0.55$. }
\label{fig:forecast_de}
\end{center}
\end{figure}

Reconstructions of $\wbg$ and $\wgr$ provide a test of dark energy and gravity. \autoref{fig:forecast_de} illustrates the potential diagnostics of this probe, using mock data sets for two non-standard models:
\\
$\bullet$~ \underline{$w_0w_a$CDM}: A dynamical dark energy model with $w(z)= w_0-w_a\,z/(1+z)$ \cite{Chevallier:2000qy,Linder:2002et}, whose fiducial values are constrained by the combination of CMB, BAO, RSD and weak lensing (last column of Table 6 in \cite{Planck:2018vyg}). \autoref{fig:forecast_de} top panel shows predictions for $\wbg$ and $\wgr$. These recover the fiducial trend well and the previous findings on precision apply. At lower $z$ the multi-task error on $\wgr$ is lower than the single-task error, which in turn is lower than the $\wbg$ error. $\wbg$ excludes the standard model at more than 95\% CL and $\wgr$ excludes it with much higher significance.
\\
$\bullet$~ \underline{$\gamma$CDM}: A simplified modified gravity model which mimics a $\Lambda$CDM expansion rate but has different growth rate, where we use the parametrisation $f=\om^\gamma$ \cite{Lahav1991, Linder2005}. We choose a $-10\%$ variation of the standard value $\gamma_{\Lambda{\rm CDM}} = 0.55$, and use the code MGCLASSII \cite{Sakr:2021ylx} to compute the fiducial predictions of $f$ and $\sigma_8$. The bottom panel of \autoref{fig:forecast_de} displays the $\wbg$ and $\wgr$ predictions. $\wbg$ is fully consistent with the standard model, while $\wgr$ (ST and MT) correctly follows the prediction of the modified fiducial. In fact, at low $z$ the growth and background reconstructions exclude each other at almost 95\% CL. This highlights that there is not only valuable phenomenological information to be gained from comparing $\wbg$ and $\wgr$ predictions with the standard model, but also from comparing them with one another. 

\subsection{A new diagnostic of gravity}
\vskip2mm

We can turn the results in \autoref{fig:forecast_de} into an effective diagnostic of gravity, based on the separation between pure-background reconstruction ($H$ from cosmic chronometers in our analysis) and background + perturbation reconstruction ($f$ in our case). By comparing the reconstructions we can schematise the diagnostic as follows:
\begin{itemize}
   \item If $\wbg \neq -1$, this indicates a departure from standard $\Lambda$CDM. By \autoref{eq:wbg}, this could be caused by spatial curvature, dynamical dark energy or modified gravity (or a combination).
    \item If ${\wgr} \neq -1$, there are 3 possible outcomes: 
        \begin{itemize}
            \item[*] $\wbg = \wgr \neq -1$, so that background and growth are consistent, hinting at dynamical dark energy within General Relativity. Both reconstructions can be used to track down the redshift evolution of $w$, allowing us to test this.
            \item[*] $\wbg \neq \wgr$, so the standard $\Lambda$CDM again breaks down, but with inconsistency between background and growth. This could hint at clustering dark energy or modified gravity. 
            \item[*] $\wbg = -1$, so the background is compatible with standard $\Lambda$CDM but perturbations are incompatible. This could arise from scale-dependent growth in clustering dark energy or modified gravity models.
        \end{itemize}
\end{itemize}

\section{Conclusion}\label{sec:discussion}

We propose a new way to reconstruct the equation of state of dark energy $w$ by using the growth of large-scale structure. This approach is based on the evolution equation for the growth rate $f$. (Note that \cite{Ruiz-Zapatero:2022zpx} use the same equation but for a different purpose, i.e. to reconstruct the evolution of $\om$. For us, $\om$ is derived via reconstruction of $H$ from data.) 

There are two ways to reconstruct $w$: purely from background data ($\wbg$); adding perturbation data to background data ($\wgr$). We apply both approaches to current data, using cosmic chronometers and RSD. We find that the data is too noisy for the reconstructions of $H$ and $f$ to be robust enough under changes of GP kernels. Consequently, reconstructed $\wbg$ and $\wgr$ also show a dependency on the kernel. The method is also sensitive to an additional input: the choice of prior on $\omo$. Nevertheless, we find the reconstructed $\wbg$ and $\wgr$ to be compatible with the standard model within $2\sigma$. Slight tendencies are noticeable: suppressed growth relative to the standard model, $\wgr$ different from $\wbg$, and $\wgr > -1$. This could be a hint of deviations from the standard model that will be crucial to explore with future data.

Consequently, we explore the performance of our reconstruction technique with mock data from nominal future surveys. With enhanced precision the GP reconstructions are robust: we find that at $z=0$ the error on $\wbg$ is decreased by a factor of almost 2, while for $\wgr$ the decrease is more than a factor of 5. This implies that the forecast error on $\wgr$ is $44\%$ more precise than that on $\wbg$ at $z=0$. 

The precision on $\wgr$ can be increased further by $67\%$, using a multi-task GP reconstruction of the data -- where $f$, $\sigma_8$ and $f\sigma_8$ are reconstructed simultaneously. On the other hand, $\wgr$ precision is more degraded than $\wbg$ at higher redshifts by the uncertainties on the $\om$ reconstruction.

We also use forecast data with two non-standard models as fiducial, which are complementary: $w_0w_a$CDM, where only the background deviates; $\gamma$CDM where only the perturbations deviate ($f=\Omega_{\rm m}^\gamma$ and $\gamma\neq0.55$). In both cases, we find that the reconstructions are precise enough to detect these deviations from the standard model. This allowed us to present a diagnostic to probe the phenomenology that can cause these deviations.

Summarising, our new method can provide some fresh insight into the dark sector of the Universe, based on a GP reconstruction of the data. Our results hold within the range of validity of the GP method. Although this method is not completely model independent (see e.g. \cite{Perenon:2021uom,OColgain:2021pyh} for further discussion), it is `agnostic', in the sense that it does not require a choice of a cosmological model. Our method is able to identify the contributions to $w$ from background alone and from background + perturbations. In this sense, it is complementary and orthogonal to more standard, model-dependent data analyses techniques, with which it can be compared when future data becomes available.





\section*{Acknowledgements}

We thank Sambatra Andrianomena, St\'ephane Ili\'c and Michelle Lochner for useful discussions. LP and RM are supported by the South African Radio Astronomy Observatory and the National Research Foundation (Grant No. 75415). MM acknowledges funding by the Agenzia Spaziale Italiana (ASI) under agreement no. 2018-23-HH.0. SC acknowledges support from the Italian Ministry of Education, University and Research (\textsc{miur}) through the `Departments of Excellence 2018-2022' Grant (L.\ 232/2016) and from the Ministero degli Affari Esteri della Cooperazione Internazionale (\textsc{maeci}) – Direzione Generale per la Promozione del Sistema Paese Progetto di Grande Rilevanza ZA18GR02. CC is supported by the United Kingdom Science \& Technology Facilities Council, Consolidated Grant No.\ ST/P000592/1.

\bibliographystyle{elsarticle-num-names}
\bibliography{References}
\end{document}